\documentclass[float,epsfig,usenatbib]{mn2e}

\usepackage{graphicx}
\usepackage{appendix}
\usepackage{psfrag}
\usepackage{amsmath,amssymb}
\usepackage{threeparttable}
\usepackage{float}
\usepackage{subfigure}
\usepackage{afterpage}
\def\aaps{A\&AS}
\def\apj{ApJ}
\def\apjl{ApJL}
\def\mnras{MNRAS}
\def\pasp{PASP}
\def\aj{AJ}
\def\araa{ARA\&A}
\def\aap{A\&A}
\def\apjs{ApJS}
\def\nat{Nature}

\title[Transition galaxies at $z\sim$0]{Evolutionary paths to and from the red sequence: \\
Star formation and \hi~properties of transition galaxies at $z\sim$0}
\author[L. Cortese \& T.M. Hughes]
{L. Cortese\thanks{luca.cortese@astro.cf.ac.uk} and T.M. Hughes
\\
School of Physics and Astronomy, Cardiff University, Queens Buildings, The Parade, Cardiff, CF24 3AA, UK\\
}

\date{Accepted 2009 August 14. Received 2009 August 13; in original form 2009 June 22}

\begin{document}
\newcommand{\Zsolar}{\mbox{$\,\rm Z_{\odot}$}}
\newcommand{\Msolar}{\mbox{$\,\rm M_{\odot}$}}
\newcommand{\Lsolar}{\mbox{$\,\rm L_{\odot}$}}
\newcommand{\xs}{$\chi^{2}$}
\newcommand{\dxs}{$\Delta\chi^{2}$}
\newcommand{\xsn}{$\chi^{2}_{\nu}$}
\newcommand{\ls}{{\tiny \( \stackrel{<}{\sim}\)}}
\newcommand{\gs}{{\tiny \( \stackrel{>}{\sim}\)}}
\newcommand{\asec}{$^{\prime\prime}$}
\newcommand{\amin}{$^{\prime}$}
\newcommand{\mstar}{\mbox{$M_{*}$}}
\newcommand{\hi}{H{\sc i}}
\newcommand{\hii}{H{\sc ii}\ }
\newcommand{\kms}{km~s$^{-1}$\ }

\maketitle

\label{firstpage}

\begin{abstract}
We investigate the properties of galaxies between the 
blue and the red sequence (i.e., the transition region, 4.5$<NUV-H<$6 mag) by combining ultraviolet (UV) and 
near-infrared imaging to 21 cm \hi~line observations for a volume-limited sample of nearby galaxies.
We confirm the existence of a tight relation between colour and \hi-fraction across all 
the range of colours, although outside the blue cloud this trend becomes gradually weaker.
Transition galaxies are divided into two different families, according to their atomic hydrogen content.
`\hi-deficient' galaxies are the majority of transition galaxies in our sample. 
They are found in high density environments and all their properties are consistent with 
a quenching of the star formation via gas stripping. 
However, while the migration from the blue cloud is relatively quick (i.e., \ls1 Gyr), 
a longer amount of time (a few Gyr at least) seems required to completely suppress the star 
formation and reach the red sequence.
At all masses, migrating `\hi-deficient' galaxies are mainly disks, implying that the 
mechanism responsible for today's migration in clusters cannot have played a significant 
role in the creation of the red sequence at high-redshift. 
Conversely, `\hi-normal' transition galaxies are a more heterogeneous population.
A significant fraction of these objects show star formation 
in ring-like structures and evidence for accretion/minor-merging events suggesting that 
at least part of the \hi~reservoir has an external origin.
The detailed evolution of such systems is still unclear, but our analysis suggests that, 
in at least two cases, galaxies might have migrated back from the red sequence after accretion events.
Interestingly, the \hi~available may be sufficient to sustain star formation at the current 
rate for several billion years.
Our study clearly shows the variety of evolutionary paths leading to the transition 
region and suggests that the transition galaxies may not be always associated with 
systems quickly migrating from the blue to the red sequence.

\end{abstract}


\section{Introduction}
The last decades have seen the rise and success of the hierarchical paradigm for 
galaxy formation in a cold dark-matter dominated universe.
Although very powerful, the concordance model is still far from providing 
us with a complete and coherent view of how galaxies form and evolve.
This is mainly because we still do not understand the physics involving 
the baryonic component. 
The current challenge for galaxy formation and evolution studies is thus to 
improve our knowledge of the astrophysical processes responsible for 
{\it transforming} simple dark matter halos into the bimodal population of 
galaxies inhabiting today's universe.

It is in fact well established that, when we look at their integrated optical colours, galaxies 
constitute a bimodal population (e.g., \citealp{tully82,baldry04}) composed of  
a `red sequence', dominated by old stellar populations, and a `blue cloud' 
where the vast majority of new stars in the universe are formed. 
However, the dichotomy in the colour distribution does not automatically reflect 
a difference in morphological type (e.g., light distribution) and we now know that 
the red sequence is not only composed of quiescent early-type galaxies (e.g., \citealp{scodeggio02,franzetti07}). 
For example, while the red sequence accounts for $\sim$60-85\% 
(depending on the colour cut used to define star-forming galaxies) of the total stellar-mass density in the 
local universe (e.g., \citealp{baldry04,bell03,borch06,perez08}), 
stars in late-type galaxies contribute to at least half the local stellar mass budget 
(e.g., \citealp{driver06,driver07b,driver07a,kochanek01}).  
This automatically implies that a significant fraction of massive late-type galaxies lie in 
the red sequence, whereas high-mass blue ellipticals are extremely rare.
Moreover, not all red galaxies have stopped forming stars, as revealed by recent ultraviolet (UV) 
investigations (e.g., \citealp{kaviraj07}). 
It thus emerges that, at least at optical wavelength, the red sequence is a heterogeneous 
family of objects which have likely followed different evolutionary paths.

How galaxies end-up in the red sequence is still partly a mystery, but recently high-redshift surveys 
have made it possible to start tracing the growth of the star-forming and quiescent galaxy population 
up to $z\sim$1 and beyond. 
Despite the observational and theoretical uncertainties (e.g., \citealp{conroy08}), it 
seems now commonly accepted that the stellar mass of the blue cloud shows very little growth since $z\sim$1, 
while the red sequence has grown by at least a factor $\sim$2 (e.g. \citealp{cimatti06,arnouts07,bell07,brown07,faber07}).
The most popular scenario invoked to explain the growth of red galaxies is 
a migration of a significant fraction of star-forming systems from the blue cloud. 
Although not always supported \citep{blanton06}, the possibility of an exchange of galaxies 
between the two sequences is exciting, and several theoretical and observational studies 
have started to look for the possible astrophysical processes responsible for such transition.
Several mechanisms have been proposed so far, among the most popular are 
different modes of gas accretion \citep{keres05,dekel06}, feedback from active galactic nuclei (AGN, \citealp{schawinsky09}), 
and environmental effects (e.g., \citealp{hughes09}, hereafter HC09). 
However, whether a population of migrating galaxies does really exist and what causes the quenching 
of their star formation is still not clear. 

In this context, the advent of the {\it Galaxy Evolution Explorer} (GALEX) large-area UV surveys 
is allowing us to tackle this problem from a different angle. 
Thanks to its high sensitivity to low-level star formation activity, UV magnitudes can be used 
to better discriminate between quiescent and still active (although optically-red) galaxies.
In fact, contrary to what is observed at optical wavelengths, the UV-optical colour distribution at a given 
mass is not well fitted by two gaussian distributions \citep{wyder07}, but it shows a significant 
excess of objects in the region between the blue and red sequence 
(i.e., the `so-called' transition region or `green-valley', \citealp{martin07}).
Transition galaxies may thus represent the missing link to understand if and how galaxies move 
from one population to the other.
However, it is worth reminding that, despite its potential, UV emission is significantly affected by dust 
and, only after accurate dust corrections, can the UV-optical colour be used to identify transition galaxies.
A significant fraction of galaxies found between the two sequences may in fact be composed 
of reddened systems  \citep{cowie08}.

Once transition galaxies are properly identified, a reconstruction 
of their past evolution is not straightforward. The correct discrimination between various 
physical mechanisms able to suppress star formation requires, in theory, a detailed investigation 
of {\it all} the galactic components, i.e., stars, gas and dust. 
Of particular importance is the atomic gas content (\hi), which represents the fuel for the 
future star formation activity. The mechanism responsible for the migration from the 
blue cloud must in fact inhibit the condensation of atomic into molecular hydrogen 
and the onset of star formation.
Unfortunately, not only is \hi~astronomy still technically limited to the nearby universe (e.g., \citealp{catinella08b}), but 
also our knowledge of \hi~properties of local galaxies is generally restricted to the blue cloud (e.g., \citealp{ages1367}).
The very local universe (e.g., up to the distance of the Virgo cluster) is currently the only place 
where it is possible to investigate the link between \hi-content and quenching of star formation in transition galaxies.

For all these reasons, we have collected UV to near-infrared imaging and \hi~21cm line data for a 
volume-limited sample of nearby galaxies covering different environments.
In our previous paper (HC09), we have highlighted the power of a combination 
of UV and \hi~observations to understand the properties of transition galaxies. 
Our analysis suggested a strong relationship between UV-near-infrared colour and \hi~content showing 
that migrating spirals are mainly \hi-deficient objects found in high density environments.
This result apparently rules out AGN-feedback as the main mechanism responsible for the quenching of the star 
formation in nearby spirals. 
However, a number of important questions still remain to be answered. 
Are transition galaxies really migrating from the blue to the red sequence? What are the time-scales 
of such migration? Is the quenching followed by a change in morphology? While in HC09 we have shown that 
a large fraction of spirals outside the blue cloud is \hi-deficient, this is not true for all transition spirals. 
How  have \hi-rich systems reached the transition region? 
The aim of this paper is thus to extend the analysis presented in HC09, in order to 
provide important constraints on the recent mass growth of the red sequence.

The paper is arranged as follows.
In \S~2 we briefly describe the sample and discuss possible biases related to the 
dust extinction correction. In \S~3 we define the transition region and in \S~4 discuss the 
relation between colour and gas content. The properties of transition galaxies 
are presented in \S~5 and their evolutionary histories and implications for galaxy 
evolution studies are discussed in \S~6. Finally, our main results are briefly summarized 
in \S~7.
 
Throughout the paper we use H$_{0} =\ $70 km s$^{-1}$ Mpc$^{-1}$. 
In the Virgo Cluster, where peculiar motions are dominant, we use distances as determined in \cite{gav99}.
Star formation rates (SFRs) are computed from the NUV luminosities, following the conversions 
by \cite{buat08}.

\section{The sample}
The analysis presented in this paper is based on the K-band selected sample described in HC09.
Briefly, it consists of a volume-limited sample of galaxies having 2MASS \citep{2massall} K-band magnitude 
K$_{Stot} \le$ 12 mag and distance between 15 and 25 Mpc. 
Additional selection criteria are a high galactic latitude (b $>$ +55$^{\circ}$) and low galactic extinction, 
A$_{B}$ $<$ 0.2 \citep{schlegel98}, to minimize galactic cirrus contamination.
The total sample contains 454 galaxies.  
Observations from the GALEX \citep{martin05} GR2 to GR4 data releases in the 
near- (NUV; $\lambda$=2316 \AA: $\Delta \lambda$=1069 \AA)  and far-ultraviolet 
(FUV; $\lambda$=1539 \AA: $\Delta \lambda$=442 \AA) band were available for 394 and 325 objects, 
respectively.  In the rest of the paper we focus our attention mainly on the $NUV-H$ colour, given the larger 
number statistics available.
UV magnitudes were obtained by integrating the flux over the galaxy optical size, determined at the 
surface brightness of $\mu$(B) = 25 mag arcsec$^{-2}$.
The typical uncertainty in the UV photometry is $\sim$0.1 mag and $\sim$0.15-0.20 mag in NUV and FUV respectively. 
Stellar masses are determined from H-band luminosities using the $B-V$ colour-dependent 
stellar mass-to-light ratio relation from \cite{bell03}, assuming a \cite{kroupa93} initial mass function.
Single-dish \hi~21 cm line emission data, necessary for quantifying the \hi~content of galaxies, was mainly 
taken from \cite{spring05hi}, \cite{goldmine} and the {\it NASA/IPAC Extragalactic Database} (NED).
Given the variety of sources from which the \hi~fluxes are taken, it is impossible to define 
a 21 cm sensitiveness limit for our sample. However, as discussed in \S~4, non detections 
start to be significant ($\sim$20\%) at $log(M(HI)/M_{star})\sim$-1.3 and dominate ($>$50\%) for 
$log(M(HI)/M_{star})$\ls-2.1.  
Estimates of atomic hydrogen mass or upper limits are available for $\sim$83\% (326/394) of the galaxies 
with NUV photometry.
We estimate the \hi~deficiency parameter ($DEF(HI)$) as defined by \cite{haynes}: i.e., 
the difference, in logarithmic units, between the observed \hi~mass and the value expected 
from an isolated galaxy with the same morphological type $T$ and optical linear diameter $D$:
$DEF(HI) = <\log M_{HI}(T^{obs},D^{obs}_{opt})> - log M^{obs}_{HI}$.
We used the equation in \cite{haynes} for early-type galaxies (E/S0 and earlier types) and 
the four revised values presented by \cite{solanes96} for late-types (Sa-Sab, Sb, Sbc, Sc and later) to calculate the 
expected \hi~mass from the optical diameter.
The regression line coefficients are almost identical from Sa to Sbc types, whereas they 
both significantly vary going to E/S0 or to Sc and later types \citep{haynes,solanes96}.
The typical uncertainty in the estimate of $DEF(HI)$ is $\sim$0.3 (e.g. \citealp{haynes,fumagalli09}), but 
it might slightly increase for dwarf galaxies and early-type systems. 
Given its large uncertainty, in the following, we will mainly use the \hi~deficiency to select those 
galaxies which have likely lost a significant 
amount of atomic hydrogen. In detail, we use a threshold of $DEF(HI)=$0.5 to discriminate 
between `\hi-deficient' and `\hi-normal' galaxies. \hi-deficient systems are thus objects with $\geq$70\% less atomic 
hydrogen than expected for isolated objects of the same optical size and morphological type.

\begin{figure*}
\centering
\includegraphics[width=12.5cm]{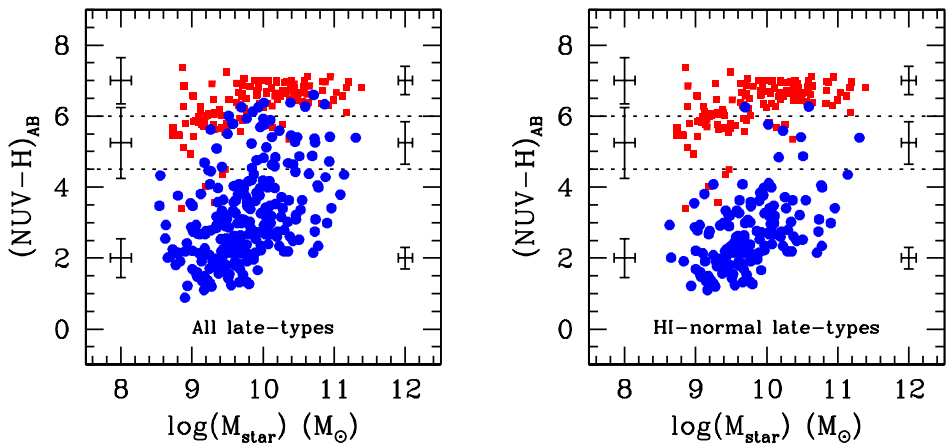}
\caption{\label{CMall} The $NUV-H$ colour-stellar-mass relations for our sample.
Colours are corrected for internal and Galactic extinction, as described in \S~2.1. 
Late and early-types are indicated with circles and squares respectively. 
The separation between blue and red sequence is clearly evident in the right panel 
where only \hi-normal late-type galaxies are shown. 
The dotted lines in both panels show the boundaries of the 
transition region as defined in \S~\ref{deftrans}. 
Typical errors, including the uncertainty in the extinction correction, are shown.}
\end{figure*}

\subsection{Extinction correction}
The estimate of the internal UV dust attenuation $A(UV)$ is a crucial step for a correct interpretation of 
the colour-mass diagram. This is particularly true outside the blue cloud where a) evolved 
stellar populations can contribute to the dust heating \citep{afuv_luca}, b) a significant fraction of the UV emission 
may come from evolved stars and not from young stellar populations (in particular in the red sequence, e.g. \citealp{bosell05}).
Previous works have shown how dramatic the effect of inaccurate dust corrections can be on the 
transition region and red sequence \citep{wyder07,schiminovich07,afuv_luca}.
\cite{afuv_luca} have recently calibrated new recipes to estimate the dust attenuation, taking into 
account the contribution of evolved stellar population to the dust heating. 
Although this method provides more realistic dust corrections with an average error of $\sim$0.5 mag in 
$A(NUV)$, the uncertainty dramatically increases up to $\sim$1 mag for galaxies lying 
between the blue and the red sequence. 
Thus, the transition region can only be investigated using a statistical approach and a comparison between 
colours and SFRs of transition galaxies could be meaningless given such large uncertainties.
 
Unfortunately, all dust correction recipes developed so far are calibrated on late-type/star-forming galaxies 
whereas in the case of early-type/quiescent systems (the dominant population in the red sequence) 
it is not a priori appropriate to apply such corrections. 
The UV emission coming from old stellar populations should in fact be less affected by dust. 
For example, a systematic (but likely unreal) shift of $\sim$0.5 mag is applied to the red sequence 
if the extinction corrections calibrated on late-types are applied to early-type galaxies \citep{afuv_luca}.
In order to avoid any systematic overestimate of the SFR in early-type galaxies (and also 
to be consistent with previous works) here we corrected all galaxies for Galactic extinction according to \cite{schlegel98}, 
but we applied internal extinction corrections only to late-type galaxies.
In detail, the internal dust attenuation was determined using the total infrared (TIR) to UV luminosity ratio method 
(e.g., \citealp{xu95}) and the age-dependent relations of \cite{afuv_luca}. 
The TIR luminosity is obtained from IRAS 60 and 100 $\mu$m fluxes or, in the few cases when 
IRAS observations are not available, using the empirical recipes described in \cite{COdust05}. 

Although this technique is likely to overestimate the colour of the few early-type galaxies 
lying in the blue cloud, we can exclude that it significantly affects the properties of transition region 
early-type galaxies and peculiar gas-rich ellipticals studied in the following sections.
To test this hypothesis, we applied the dust attenuation corrections described above to these objects finding that, 
even after correcting for dust, transition early-types still lie in the transition region and peculiar red ellipticals 
are still in the red sequence. 
However, in the following, we will try to discuss as much as possible any systematic error that could 
be introduced by problems in the extinction corrections and all error-bars shown in our figures take into 
account the uncertainty in the estimate of $A(UV)$.

\begin{figure}
\includegraphics[width=8.5cm]{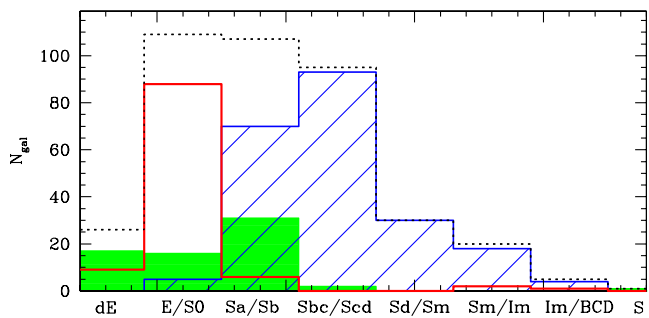}
\caption{\label{typedistr} The morphological 
type distribution of galaxies in our sample (dotted histogram). 
The filled histogram represents galaxies in the transition region, while 
red and blue sequence galaxies are shown with the empty and dashed histogram, respectively. }
\end{figure}

\section{Where does the blue cloud end?}
\label{deftrans}
Contrary to what is observed in optical \citep{baldry04},  
at fixed stellar mass, the UV-optical colour distribution of galaxies is 
not best fitted with the sum of two gaussians.
An excess of galaxies is clearly present 
in between the two sequences \citep{wyder07}. This is likely due to galaxies 
with suppressed star formation perhaps migrating from the blue to the red sequence. 
However, it is not straightforward to clearly define the range of colours 
characterizing galaxies which do not belong either to the blue cloud or the red sequence. 

On one side, it is difficult to determine 
(both observationally and theoretically) where the blue cloud ends. 
In this paper we decide to use the blue cloud of \hi-normal  
galaxies to determine the range of colours typical of unperturbed galaxies. 
This is motivated by the fact, discussed in HC09, that 
the transition region is mainly populated by \hi-deficient galaxies 
in high-density environments. 
\hi-normal late-type galaxies form a blue cloud clearly separated 
from the red sequence at all masses, as shown in the right panel of Fig.~\ref{CMall}.
We thus define galaxies with suppressed star formation as those 
objects with $NUV-H>$4.5 mag (corresponding to the 90th percentile of the colour 
of \hi-normal spirals with M$_{star}>$10$^{10}$ M$_{\odot}$), consistently with \cite{martin07b}.
This colour-cut roughly corresponds to star formation histories (SFHs) having 
e-folding time-scale of $\sim$3 Gyr  (assuming a galaxy age of 13.7 Gyr, solar 
metallicities and the models of \cite{BC2003}) e.g., the 
typical e-folding time dividing local late- and early-type galaxies \citep{gav02}.
We note that, adopting this convention, M31 would be classified as `normal' blue-cloud 
galaxy ($NUV-H\sim$4.1 mag).
Being calibrated on massive galaxies, our colour-cut must be considered as a 
conservative upper-limit in the case of dwarf systems.   

\begin{figure*}
\centering
\includegraphics[width=12.5cm]{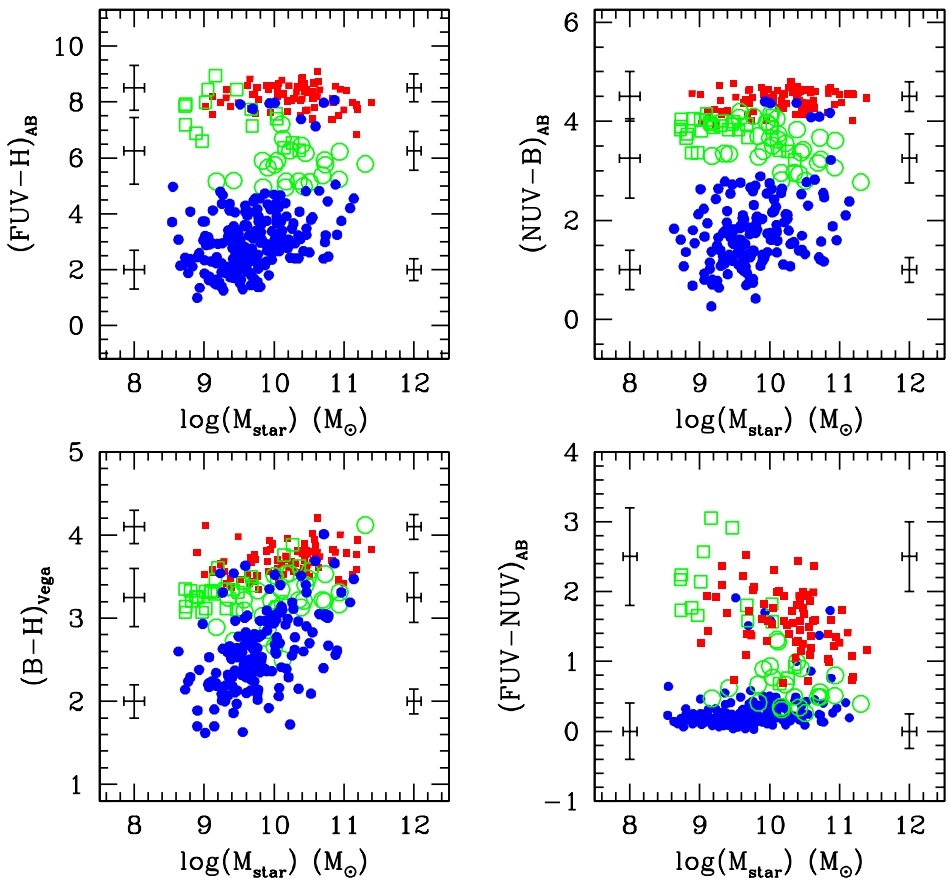}
\caption{\label{CMall2} The $FUV-H$ (upper-left), $NUV-B$ (upper-right), 
$B-H$ (lower-left) and $FUV-NUV$ (lower-right) colour-stellar-mass relations for our sample.
Symbols are as in Fig.~\ref{CMall}. Green empty symbols show transition galaxies as defined 
in \S~\ref{deftrans}. We note that, given the incomplete UV and optical 
coverage available for our sample, each graph includes a different number of galaxies.
These plots must thus be considered just as an indication of the typical colour range occupied 
by transition galaxies.}
\end{figure*}

On the other side, UV-to-near-infrared colours typical of the red sequence ($NUV-H \gtrsim$6) 
can either indicate low residual star formation activity or old, evolved stellar 
populations (e.g. \citealp{bosell05}).
The phenomenon of the UV-upturn \citep{connell} makes colours redder than $NUV-H \sim$6 
difficult to interpret so that the $NUV-H$ colour cannot be 
considered anymore as a good proxy for the specific star formation rate (SSFR). 
Following \cite{kaviraj07}, we use observations of well known 
strong UV-upturn galaxies to derive a lower limit on the $NUV-H$ colour 
typical of evolved stellar populations. 
In details, given the typical colour observed in 
M87 ($NUV-H\sim$ 6.1 mag) and NGC4552 ($NUV-H\sim$ 6.4 mag), we assumed 
$NUV-H$=6 mag as a conservative lower limit to discriminate between 
residual star formation and UV-upturn.
The validity of this colour-cut is confirmed by a visual inspection of GALEX colour 
images, which indicates that only 6\% (i.e. 6 objects) of galaxies redder 
than $NUV-H$=6 mag show clear evidence of residual SF (e.g., blue star-forming knots).

The morphological type distributions for galaxies belonging to the three 
groups here considered (i.e., blue cloud, red sequence and transition region) are 
shown in Fig.~\ref{typedistr}. It clearly emerges that red and blue galaxies are 
two disjoint families not only in colour, but also in shape. 

Of course, the criteria described above are arbitrary and vary 
according to the colour adopted and to the stellar mass range investigated.
This can be clearly seen in Fig.~\ref{CMall2}, where the position of `transition 
galaxies' in different UV and optical colour-stellar-mass diagrams is highlighted.
Although it is indisputable that not all transition galaxies are outside 
the red sequence in a $FUV-H$ and $NUV-B$ colour diagram 
(in particular at low stellar-masses), it emerges that the definition here 
adopted is able to select a statistically 
representative sample of galaxies with suppressed star formation.
Moreover, the comparison between the top and bottom row in Fig.\ref{CMall2} highlights 
the necessity of UV colours to select fair samples of transition galaxies: e.g., a simple 
$B-H$ colour-cut would significantly contaminate our sample with star-forming 
blue-sequence and quiescent red-sequence systems.
Finally, it is interesting to note the wide range in $FUV-NUV$ colour spanned by the transition galaxies, 
suggesting different current star formation rates.

As shown in the following sections, the sample of transition galaxies 
selected using these criteria is not significantly contaminated either by 
active star-forming or quiescent galaxies erroneously classified as transition systems. 
Thus, in the rest of this paper we will refer to the colour interval 
4.5$<NUV-H<$6 mag as the `transition region'.

Finally, it is worth reminding the reader that, in all the figures presented in this paper, 
the UV-near-infrared colour is directly related to the SSFR only 
outside the red sequence. For colours redder than $NUV-H\sim$6 mag, UV magnitudes  
cannot be blindly used to quantify current SFRs and the presence of a sequence 
does not imply that all red galaxies have the same SSFR, as 
shown in the following sections.

\begin{figure*}
\includegraphics[width=17.5cm]{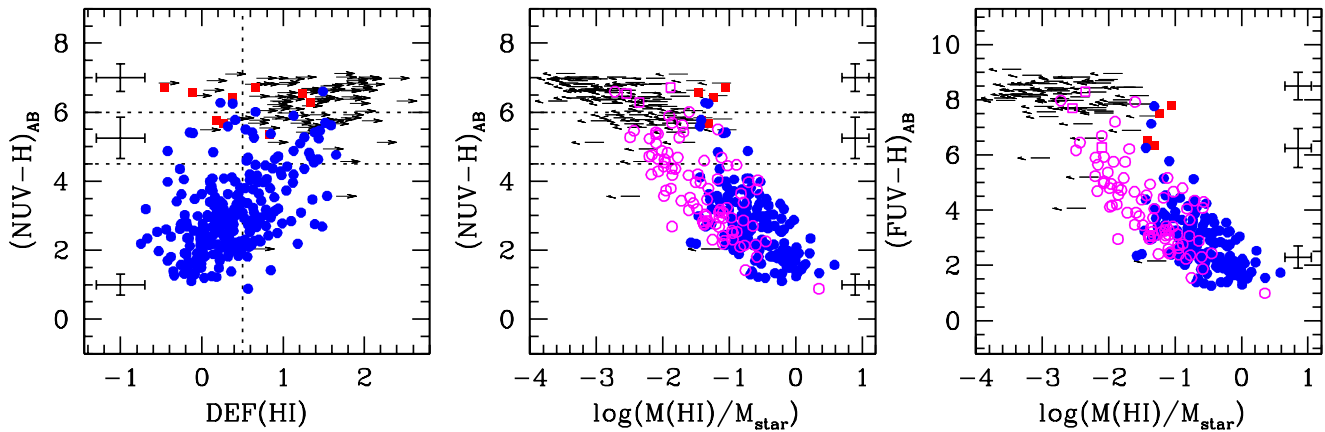}
\caption{\label{colgfrac}
The link between \hi-content and colour.
Left: $NUV-H$ colour vs. \hi~deficiency. The vertical 
dotted line separates galaxies with `normal' gas content from \hi-deficient systems. 
Center: $NUV-H$ colour vs. gas-fraction. 
Right: $FUV-H$ vs. gas-fraction. 
Late- and early- type galaxies are indicated with circles and squares respectively. 
Empty symbols highlight detected \hi-deficient galaxies, while arrows show upper limits. 
The horizontal dashed-lines show the boundaries of the transition region, as defined in \S~\ref{deftrans}.}
\end{figure*}

\section{The link between atomic hydrogen content and colour}
In HC09 we showed that late-type galaxies outside the blue cloud appear to have lost 
at least $\sim$70\% of their atomic hydrogen content, when compared with isolated galaxies 
of similar size and morphology. 
We commented this result as a strong evidence of a physical relation 
between the loss of gas and quenching of the star formation.
This interpretation is confirmed and reinforced in Fig.~\ref{colgfrac}.
The left panel shows the correlation between colour and \hi~deficiency: galaxies outside 
the blue cloud have not only lost a significant fraction of their 
atomic hydrogen content, but we also find a correlation between \hi~deficiency 
and $NUV-H$ colour, although with large scatter.
At least part of the scatter is due to the large uncertainty in 
the estimate of \hi~deficiency ($\sim$0.3 dex).
The same relation can be expressed in terms of gas-fraction (here defined 
as the ratio of the \hi~ to the stellar mass), as discussed by \cite{kannappan04} and 
shown in the central and right panels of Fig.~\ref{colgfrac}: lower gas-fractions 
correspond to redder colours.
The best linear fit to the relation (excluding upper-limits) is 
$log(M(HI)/M_{star})$=$-0.35\times(NUV-H)$+0.19, with a dispersion of 
$\sim$0.43 dex.
However, from the three panels in Fig.~\ref{colgfrac} it clearly emerges 
that not all transition region and red-sequence galaxies are \hi-deficient, but 
a number of systems have an amount of hydrogen typically observed in objects 
lying in the blue cloud.
This is particularly interesting if we look at the colour-gas-fraction relations 
(central and right panels). 
Outside the blue cloud, galaxies lie mainly at the two edges of the 
relation depending on whether they are \hi-deficient (empty circles) or not and 
the colour-gas-fraction relation appears more scattered\footnote{The fact that, 
for the same colour, \hi-deficient objects have a lower gas fraction is expected, since 
both quantities trace the specific amount of atomic hydrogen in a galaxy.}.
This suggests that, for $NUV-H>$4.5 mag, the gas-fraction 
is not a good proxy of the UV-optical colour anymore and vice-versa. 
The dispersion in the colour-gas-fraction relation increases from $\sim$0.35 dex (consistent with 
\citealp{kannappan04} and \citealp{cheng_gfrac09}) to $\sim$0.54 dex when we move 
from \hi-normal blue-cloud galaxies to transition and red-sequence objects. 
This is in reality a lower-limit on the real scatter increase since upper-limits are not included 
in the calculation.

The results shown in Fig.~\ref{colgfrac} are strongly suggestive 
of a) a different evolutionary path followed by \hi-deficient and \hi-normal galaxies 
outside the blue cloud and b) of a weaker link between \hi-content and colour 
than the one typically observed in star-forming galaxies.
Therefore, in order to gain additional insights on the evolution of galaxies 
in the transition region, in the following we divide transition galaxies 
into two families according to their gas content and investigate separately their 
properties.

Before investigating the detailed properties of transition galaxies, it is worth 
adding a few notes about the validity of the classification for \hi-normal transition 
systems. The low number of objects in this category and the large uncertainties in the 
estimate of gas fractions and UV dust attenuation might suggest that these are just 
random outliers, not different from the bulk of the \hi-deficient population.
Although we cannot exclude the presence of a few misclassified galaxies in both the 
\hi-deficient and \hi-normal population, it is very unlikely that 
all (and only) the \hi-normal galaxies outside the blue sequence are affected 
by a large ($>$0.5 dex) systematic underestimate of dust attenuation, and/or gas fraction.  
More importantly, the analysis presented in the next sections will clearly show that these 
two families have reached the transition region following different evolutionary paths.

\begin{figure}
\includegraphics[width=8.5cm]{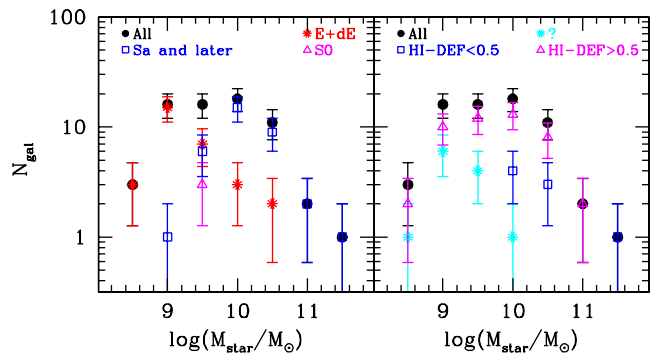}
\caption{\label{trdistr} The stellar mass distribution of transition galaxies (filled circles). 
Left: Spirals (empty squares), lenticulars (triangles) and E+dE (asterisks) are indicated.
Right: Galaxies are highlighted according to their \hi~content. Galaxies for which the estimate 
of \hi~deficiency is unsure are indicated with asterisks. }
\end{figure}

\begin{figure*}
\includegraphics[width=17.5cm]{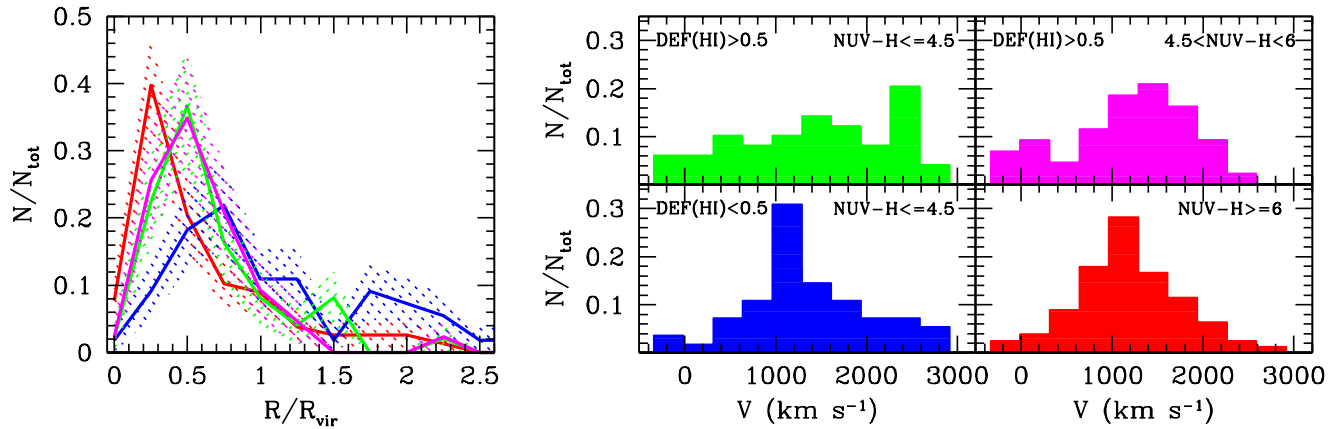}
\caption{\label{cluster} The cluster-centric projected-distance (left) and line-of-sight velocity distribution (right) of galaxies in the Virgo cluster region. 
Galaxies are divided into 4 sub-samples according to their colour and \hi-content: blue-cloud \hi-normal  
($NUV-H\leq$4.5 and $DEF($\hi$)<$0.5, blue), blue-cloud \hi-deficient ($NUV-H\leq$4.5 and $DEF($\hi$)>$0.5, green), 
transition \hi-deficient (4.5$<NUV-H<$6 and $DEF($\hi$)>$0.5, magenta) and red-sequence ($NUV-H\geq$6, red) 
galaxies. The shaded areas in the left panel show the uncertainty in the 
radial distributions. }
\end{figure*}

\section{The properties of transition galaxies}
In total, 67 galaxies in our sample, corresponding 
to $\sim$17\% in both number and total stellar-mass lie in the 
transition region as defined in \S~3.
In Fig.~\ref{trdistr}, we show the stellar-mass distribution of transition galaxies 
divided according to their morphological type (left panel) and gas content (right panel). 
For $\mstar \gtrsim 10^{10}$ \Msolar, galaxies with 4.5$<NUV-H<$6 mag are mainly spirals, 
whereas at lower stellar masses they are preferentially dwarf elliptical systems. 
More importantly, the majority of transition galaxies have $\gtrsim$70\% less 
atomic hydrogen content than isolated galaxies of similar optical size and morphological 
type. However, as already noted in HC09, not all galaxies in the transition region 
are \hi-deficient. 
For $\mstar \gtrsim 10^{10}$ \Msolar, $\sim$30\% (8 galaxies) of the transition 
galaxies have \hi~deficiency lower than 0.5. 
For lower stellar masses, it is difficult to quantify the number of gas-rich objects. 
\hi~observations are not available for 9 of our galaxies and for 3 additional 
objects the lower limits obtained for the \hi~deficiency are below our threshold of 0.5.
These are mainly dE cluster galaxies (Fig.~\ref{trdistr}, left panel), suggesting that 
their evolution is related to the cluster environment \citep{dEale}.
However, to be conservative, in the following analysis we will focus our attention on the 55 
galaxies for which the classification as \hi-deficient or \hi-normal galaxy is reliable.

\subsection{\hi-deficient systems}
\label{hidef}
Overall, sure \hi-deficient galaxies represent $\sim$70\% (47 galaxies) in number and $\sim$63\% 
in stellar mass of the transition region\footnote{This fraction might increase up to $\sim$88\% in case all galaxies 
without \hi~measurement are \hi-deficient galaxies. We note, however, that these values may not be representative of the 
local universe, being our sample likely biased towards high-density environments (see \S~6).}.
All except four galaxies lie in the Virgo cluster region (as defined in \citealp{goldmine}) suggesting 
that the cluster environment is playing an important role in quenching the star formation. 
Additional support to this scenario is obtained when we consider the properties of galaxies 
divided according to their gas content and their position in the colour-mass diagram. 
We compared the median projected distance from the cluster center
of the different populations. Given the large asymmetry of Virgo and the presence of two main 
sub-clusters (Virgo A and Virgo B, at $\sim$1 virial radii projected-distance), 
for each galaxy we computed the projected-distance from the center 
of both clouds and adopted the smallest of the two values. 
Our results do not qualitatively change if just the distance from M87 is adopted. 
The median cluster-centric distance decreases from $\sim$0.83 virial radii (R$_{vir}$), in case 
of \hi-normal blue-cloud galaxies, to 0.51, 0.52 and 0.42 R$_{vir}$ for \hi-deficient blue-cloud, 
\hi-deficient transition and red-sequence objects, respectively (see Fig.~\ref{cluster}, left panel).
Similarly, the difference between the 25th and 75th percentiles of the line-of-sight velocity distribution (i.e. 
a good estimate of the velocity dispersion in case of non gaussian distributions) 
increases in the blue cloud from $\sim$760 \kms to $\sim$1400 \kms when we consider 
\hi-normal and \hi-deficient galaxies respectively. Then, the typical velocity dispersion 
gradually decreases to $\sim$915 \kms and $\sim$610 \kms when we consider the transition region 
and the red sequence respectively (see Fig.~\ref{cluster}, right panel). 
The gradual variation in projected distance and velocity distribution when moving in the colour magnitude diagram 
from blue, \hi~normal systems to red, quiescent objects supports the idea that \hi-deficient 
galaxies represent a population of galaxies recently infalled into the cluster and not yet virialized.
For example, a free-falling population is expected to have a velocity dispersion $\sqrt[]{2}$ times larger 
than the virialized population. 
Moreover, the velocity dispersion profile for \hi-deficient galaxies decreases with 
cluster-centric distance consistent with isotropic velocities in the center and radial 
velocities in the external regions, 
as expected in the case of galaxy infall onto the cluster \citep{GIRARD98,COGA04}.
The opposite trend (i.e., increasing with cluster-centric distance) is observed for 
red-sequence galaxies, 
as expected in a relaxed cluster undergoing two-body relaxation in the dense central region, 
with circular orbits in the center and more isotropic velocities in the external regions.
Finally, a visual investigation of UV images reveals that in at least 50\% of star-forming Virgo 
galaxies in the transition region the star formation is only present well within the optical radius,  
completing the collection of evidence supporting environmental effects behind the quenching 
of the star formation in \hi-deficient Virgo galaxies.

Less clear is the origin of the \hi~deficiency in galaxies outside Virgo: 4 objects in total, 
namely NGC4684, UGC8756, UGC8032 and NGC5566.
NGC5566 is the brightest member of a galaxy triplet while UGC8032 lies just $\sim$1.1 virial radii from 
the center of Virgo. Thus for these two galaxies it is still possible that 
environmental effects are playing a role in the gas stripping. 
The fact that UGC8032 is not included in our Virgo sample despite its small distance from M87 is 
due to the fact that it just lies outside the Virgo boundaries defined by \cite{goldmine}. 
The origin of the \hi~deficiency in NGC4684 and UGC8756 remains a puzzle. 
UGC8756 has in fact no nearby companions or any clear sign of interaction. 
NGC4684 is a lenticular galaxy with very strong UV nuclear emission, probably related to the extended H$\alpha$ outflow 
discovered by \cite{bettoni93}. The outflow has been interpreted as related to bar instability and it is not 
clear whether such process can be responsible for the \hi~deficiency observed in this object.
Thus, the origin of the \hi~deficiency in these two objects still remains unclear.

\subsubsection{Time-scales for the migration}
Combining the observational evidence presented above with previous works on the Virgo cluster 
(e.g. \citealp{review} and references therein), 
we favor hydrodynamical interactions like ram-pressure as the main process responsible for the suppression 
of the star formation. However, we note that gravitational interactions cannot be excluded in at least one case 
(NGC4438, e.g., \citealp{kenney08,n4438,vollmer05}). 

\begin{figure}
\centering
\includegraphics[width=6.cm]{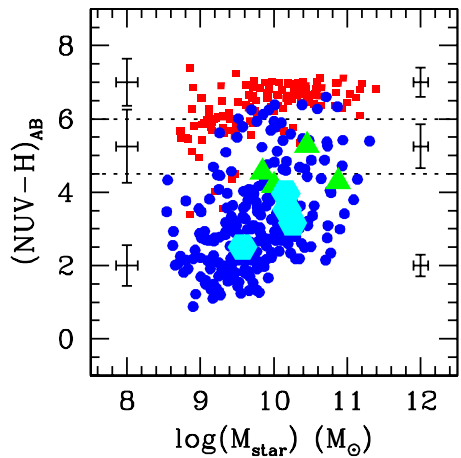}
\caption{\label{timescale} Same as Fig.~\ref{CMall} (left panel). Large symbols indicates 
galaxies in the Crowl \& Kenney (2008) sample for which a stripping time-scale estimate is available.
Stripping time-scale shorter than 300 Myr and between 300-500 Myr are shown with 
hexagons and triangles respectively. }
\end{figure}

Only recently, has it become possible to accurately quantify the time-scale for the quenching of the 
star formation after the stripping event \citep{n4569,crowl2008}. 
Luckily, almost all the objects for which a stripping time-scale has been computed are included in our sample.
In Fig.~\ref{timescale}, we highlight the position of these six objects in the colour-mass diagram according to the 
age of the stripping event: hexagons and triangles indicate galaxies in which the star formation in the outer regions 
has been suppressed less or more than $\sim$300 Myr, respectively. 
Interestingly, there is a difference in the average colour between very recent quenching ($t<$300 Myr) and older events, 
so that only galaxies with quenching time-scale $\sim$400-500 Myr are at the edge or have already reached the 
transition region. 
This is consistent with the fact that the Virgo transition galaxy population is not virialized, implying a recent 
($\leq$ 1.7 Gyr, i.e., the Virgo crossing time; \citealp{review}) infall into the cluster center. 
Thus, we can conclude that once the \hi~has been stripped from the disk, a galaxy moves from the blue cloud 
to the transition region in a time-scale roughly $\sim$0.5-1 Gyr.

It is more difficult to predict the future evolution of transition galaxies and in particular 
to estimate how long they will remain in the transition region and whether they will eventually 
join red-sequence galaxies.    
As already pointed out by \cite{crowl2008} and \cite{dEale}, one cluster-crossing is not sufficient to 
completely halt the star formation in massive (M$_{star}$\gs 10$^{10}$ M$_{\odot}$) galaxies. 
In fact, while the outer disk is completely deprived of its gas content and 
star formation is quickly stopped, in the central regions the restoring force is too strong, keeping 
the atomic hydrogen reservoir necessary to sustain continuos star formation.
Moreover, \cite{dEale} showed that two cluster crossings ($\sim$2-3 Gyr) are already necessary to move the brightest 
dE from the blue to the red sequence, suggesting that the transition galaxies described here will take 
at least the same amount of time to have their star formation completely quenched.
Assuming that the \cite{GUNG72} formalism for ram-pressure is still valid after the first passage and that 
the galaxy's orbit does not change significantly, we can expect that very little additional gas will be stripped 
during the second passage by ram-pressure. 
Significant stripping would occur only if the galaxy's restoring force is lowered by gravitational interactions 
with other members and the cluster potential well. 
Other intra-cluster medium related environmental effects, like viscous stripping \citep{NUL82} and 
thermal evaporation \citep{COWS77}, may thus play an important role in the complete suppression of the star formation. 

An upper limit to the time spent by \hi-deficient galaxies in the transition region can be obtained if 
we assume that all the remaining gas will be consumed by star formation. 
It is in fact plausible that the intra-cluster medium will prevent additional infall of cold gas.
In this case, the `Roberts' time' (\citealp{roberts63}, defined as the ratio of the gas mass to the current star formation 
rate: i.e., M(gas)/SFR) can be used to obtain a rough estimate of the gas consumption time. 
Assuming that 15\% of the total gas is in the molecular state \citep{boselligdust} and $\sim$30\% is composed 
by helium and heavy elements \citep{boselli}, we find that the `Roberts' time' is already $\sim$2.2 Gyr\footnote{This 
value decreases by a factor $\sim$1.5 if a Salpeter IMF is adopted}.
This is in reality a lower limit to the real value since it does not take into account gas recycling.
As shown by  \cite{kennbirth}, the real gas consumption time is 1.5-4 times longer then the time scale 
calculated above. Thus, although they have lost a significant amount of their original gas content, \hi-deficient 
transition galaxies still have enough fuel to sustain star formation at the current rate for at least a couple of Gyr.

Detailed simulations focused on the effect of the cluster environment after the first passage will thus be extremely 
interesting to understand the future evolution of these systems. 
At this stage, the main conclusion we can draw from our analysis is that 
at least $\sim$3 Gyr seem to be necessary for the complete migration of a galaxy from the blue to the 
red sequence when gas stripping via the intra-cluster medium is involved.

\begin{table*}
\scriptsize
\caption{The properties of \hi-normal transition and red-sequence galaxies in our sample.}
\label{hirich}
\begin{tabular}{l@{\hspace{0.2pt}}c@{\hspace{8pt}}c@{\hspace{8pt}}c@{\hspace{8pt}}c@{\hspace{8pt}}c@{\hspace{8pt}}c@{\hspace{8pt}}c@{\hspace{8pt}}c@{\hspace{6pt}}c@{\hspace{6pt}}c@{\hspace{4pt}}c@{\hspace{6pt}}c@{\hspace{6pt}}c@{\hspace{6pt}}c@{}}
\hline
\noalign{\smallskip}
\multicolumn{15}{c}{\rm \hi -normal~transition~galaxies}\\
\hline
NAME & TYPE   &   D      &AGN   &  B-V  &  FUV	 & NUV	 & H   & F$_{60 \mu m}$ & F$_{100 \mu m}$ & M(HI) & M(HI)/M$_{star}$ & C$_{31}$(H) & Merging/Accretion? & Ref. \\
     &        &  Mpc     &      &       & m$_{AB}$ & m$_{AB}$  & m$_{AB}$ & Jy  & Jy              & 10$^{8}$ M$_{\odot}$ &   &              &                 &  \\ 
\noalign{\smallskip}
\hline
\noalign{\smallskip}	 
NGC3619 & S0 & 20.7 & 	-  &  0.86& 16.86  & 16.14  & 7.40   & 0.43 & 1.61  & 7.08  & 0.04 & 6.2 &   yes & 1,2 \\
NGC3898 & Sa & 15.7 & 	Lin&  0.79&  -	   & 15.03  & 6.25   & 0.42 & 2.02  & 26.9  & 0.09 & 3.1 &   may~be & 3 \\
NGC4324 & Sa & 17.0 & 	Lin/Sey&  0.87& 16.62 & 15.99 & 7.29 & 0.45 & 1.96  & 6.76  & 0.05 & 3.5 &   - & - \\
NGC4370 & Sa & 23.0 & 	NoL  &  0.70& -      & 18.26  &8.28  & 0.94 & 3.27  & 4.0   & 0.04 & 3.8 &   yes  & 4,5 \\
NGC4378 & Sa & 17.0 & 	Sey&  0.81& 16.00  & 15.50  & 7.23   & 0.36 & 1.45  & 10.0  & 0.07 & 4.9 &   - & - \\
NGC4772 & Sa & 17.0 & 	Lin&  0.87& 17.07  & 16.16  & 7.12   & 0.38 & 1.32  & 6.17  & 0.04 & 5.1 &   yes & 6 \\
NGC5701 & Sa & 20.1 & 	Lin&  0.84& 15.48  & 15.11  & 6.74   & 0.27 & 1.36  & 61.7  & 0.20 & 4.0 &   - & - \\
\noalign{\smallskip}									
\hline
\noalign{\smallskip}									
\multicolumn{15}{c}{\rm \hi -normal~ red-sequence~ galaxies}\\
\hline
NAME & TYPE   &   D      &AGN   &  B-V  &  FUV	 & NUV	 & H   & F$_{60 \mu m}$ & F$_{100 \mu m}$ & M(HI) & M(HI)/M$_{star}$ & C$_{31}$(H) & Merging/Accretion? & Ref. \\
     &        &  Mpc     &      &       & m$_{AB}$ & m$_{AB}$  & m$_{AB}$ & Jy  & Jy              & 10$^{8}$ M$_{\odot}$ &   &             &                 &  \\ 
\noalign{\smallskip}
\hline
\noalign{\smallskip}										
NGC4203       & S0  &  17.0 &  Lin. &  0.99 &16.94	 & 15.87  & 6.26   & 0.59 & 2.16 & 33.1 & 0.09 & 6.8 &  yes & 7 \\
NGC4262$^{a}$ & S0  &  17.0 &  NoL   &  0.83 &-         & 16.9   & 7.20   & -    & 0.50 & 5.13 & 0.04 & 6.0  &  yes & 8 \\
NGC4698$^{a}$ & Sa  &  17.0 &  Sey &  0.83 &16.64	 & 15.71  & 6.16   & 0.63 & 1.89 & 17.0 & 0.04 & 4.2 &  yes  & 9 \\
NGC4866       & S0  &  17.0 &  Lin &  0.96 &17.26	 & 16.27  & 6.76   & -    & -	 & 13.5 & 0.06 & 3.7 &  - & - \\
NGC5103       & Sab &  17.0 &  -   &  -    &19.49	 & 17.76  & 8.29   & -    & -	 & 2.40 & 0.05 & 6.3 &  - & - \\
\noalign{\smallskip}
\hline
\end{tabular}
\begin{flushleft}
$^{a}$ UV fluxes increase significantly if the UV rings outside the optical radius are included.\\
References: (1) \cite{vandriel89}; (2) \cite{howell06}; (3) \cite{noordermeer05}; (4) \cite{bertola88}; (5) \cite{patil09}; (6) \cite{haynes00}; 
(7) \cite{vandriel88}; (8) \cite{krumm85}; (9) \cite{bertola99} \\
\end{flushleft}
\end{table*}

\begin{figure*}
\includegraphics[width=17.5cm]{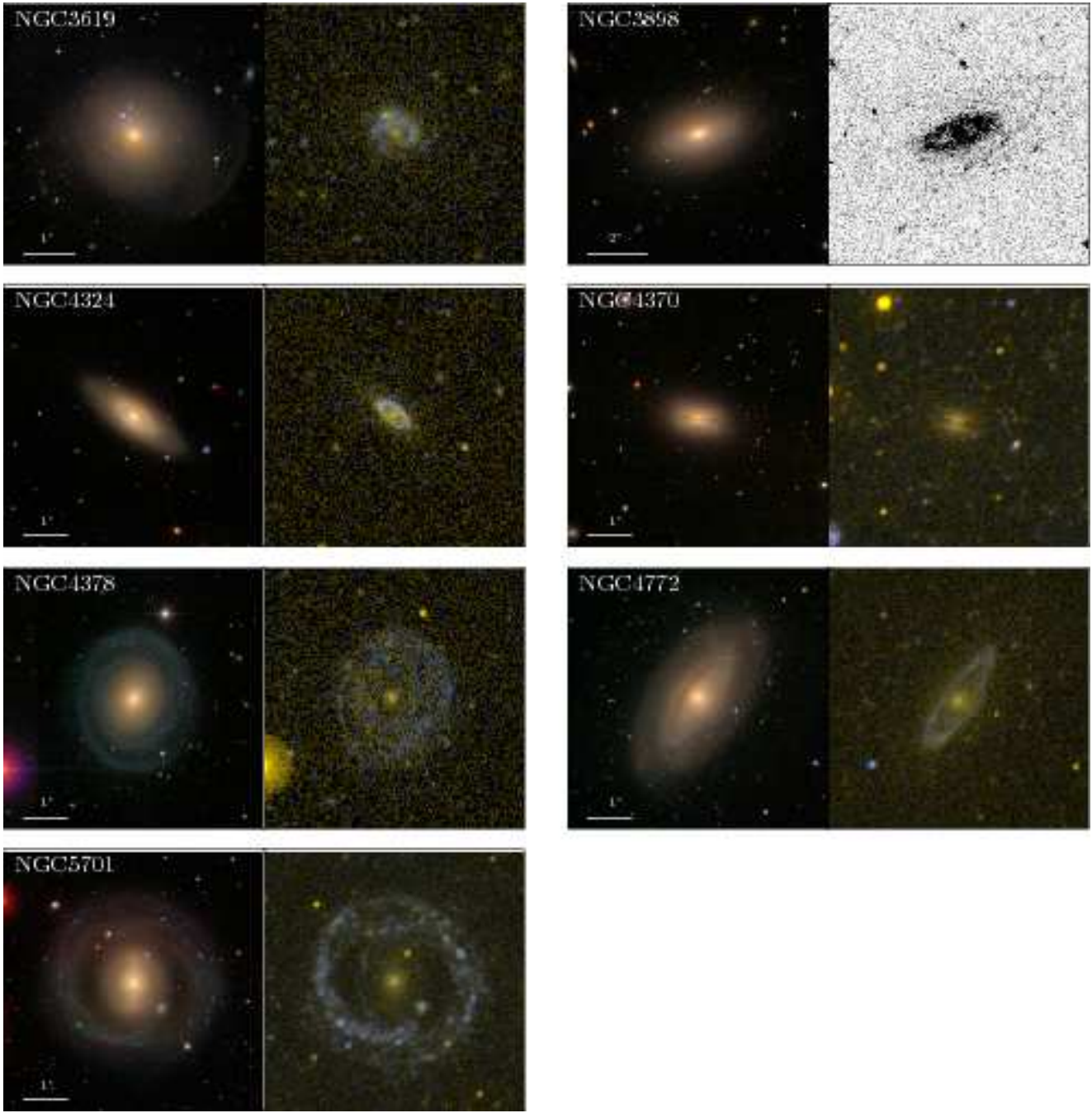}
\caption{\label{FCtrans} \hi-normal transition galaxies. For each galaxy, the SDSS RGB and GALEX 
FUV-NUV colour images are shown. In case GALEX FUV observations are not available the NUV image is shown in 
black and white.}
\end{figure*}

\subsection{\hi-normal galaxies}
Contrary to \hi-deficient objects, sure \hi-normal transition galaxies are equally distributed 
between the field and cluster environments.
They are $\sim$12 \% in number\footnote{This fraction might increase up to $\sim$30\% in case all galaxies 
without \hi~measurement are \hi-normal galaxies.} (8 objects in total) 
and $\sim$34\% in stellar mass of the whole transition galaxy population in our sample.
We note that NGC4565 is the only example of perfectly edge-on transition object, 
in this case we cannot exclude that the corrections adopted still underestimate the real UV extinction  \citep{panuzzo03} 
making this galaxy possibly an erroneous transition object. Therefore, we will exclude NGC4565 from the following analysis.
The properties of the remaining 7 objects are listed in Table~\ref{hirich} as follows:
Col. 1: Name. Col.2: Morphological type. Col. 3: Distance in Mpc. Col.4 AGN classification (following 
the criteria described in \citealp{decarli07}): Lin=LINER, Sey=Seyfert, NoL= No emission lines. Col.5: $B-V$ colour 
corrected for dust extinction. Col.6-8: FUV, NUV and H AB magnitudes. Col.9-10: IRAS fluxes at 60 and 100 $\mu m$. 
Col. 11: HI mass. Col. 12: \hi- to stellar-mass ratio. Col. 13: concentration index in H band taken from 2MASS ($C_{31}(H)$ 
defined as the ratio between the radii containing 75\% and 25\% of the total H-band light). Col. 
14-15: Note regarding any evidence (and relative reference) 
supporting an external origin for the \hi.
In Fig.~\ref{FCtrans}, we show SDSS optical and GALEX UV colour images for each galaxy.
In Appendix~A, we describe the properties of each object in 
order to investigate the possible origins for the normal \hi~content and low SSFR.
From this analysis, it emerges that \hi-normal transition galaxies are a heterogeneous 
class of objects going from merger remnants (e.g., NGC3619) 
with star-formation activity limited to the center, to satellites of big ellipticals (NGC4370), 
with no evident signs of recent star formation. 
Contrary to the \hi-deficient family, gas stripping by environmental effect seems not to be playing any role 
in their recent evolution.
As expected in the case of massive ($M_{star}\geq10^{10}$ M$_{\odot}$) galaxies, the majority of 
these objects host an `optical' AGN.
More surprisingly, our analysis shows that, although they have likely followed 
different evolutionary paths, a significant fraction of galaxies (at least 4 out of 7) has  
recent star formation mainly in form of one or more UV rings. 
As shown in Fig.~\ref{FCtrans}, the UV rings have different morphologies going from inner 
rings (NGC3898, NGC4324, NGC4772) to Hoag-like objects (NGC5701).
Resolved \hi~maps, available for two of our objects (NGC3898, NGC4772), reveal that the \hi~is 
distributed in extended low surface density disks, exceeding significantly the typical column density 
of 1-2 M$_{\odot}$ pc$^{-2}$ only in correspondence of the star forming rings.
Thus, in these cases, star formation is reduced not because the \hi~has been stripped but just because 
the gas is not able to collapse into stars efficiently. 
This is likely due to the fact that the gas reservoir has a typical column density well below the critical 
density necessary to convert \hi~into molecular hydrogen and onset the star formation \citep{krumholz09}. 

\begin{figure*}
\includegraphics[width=17.5cm]{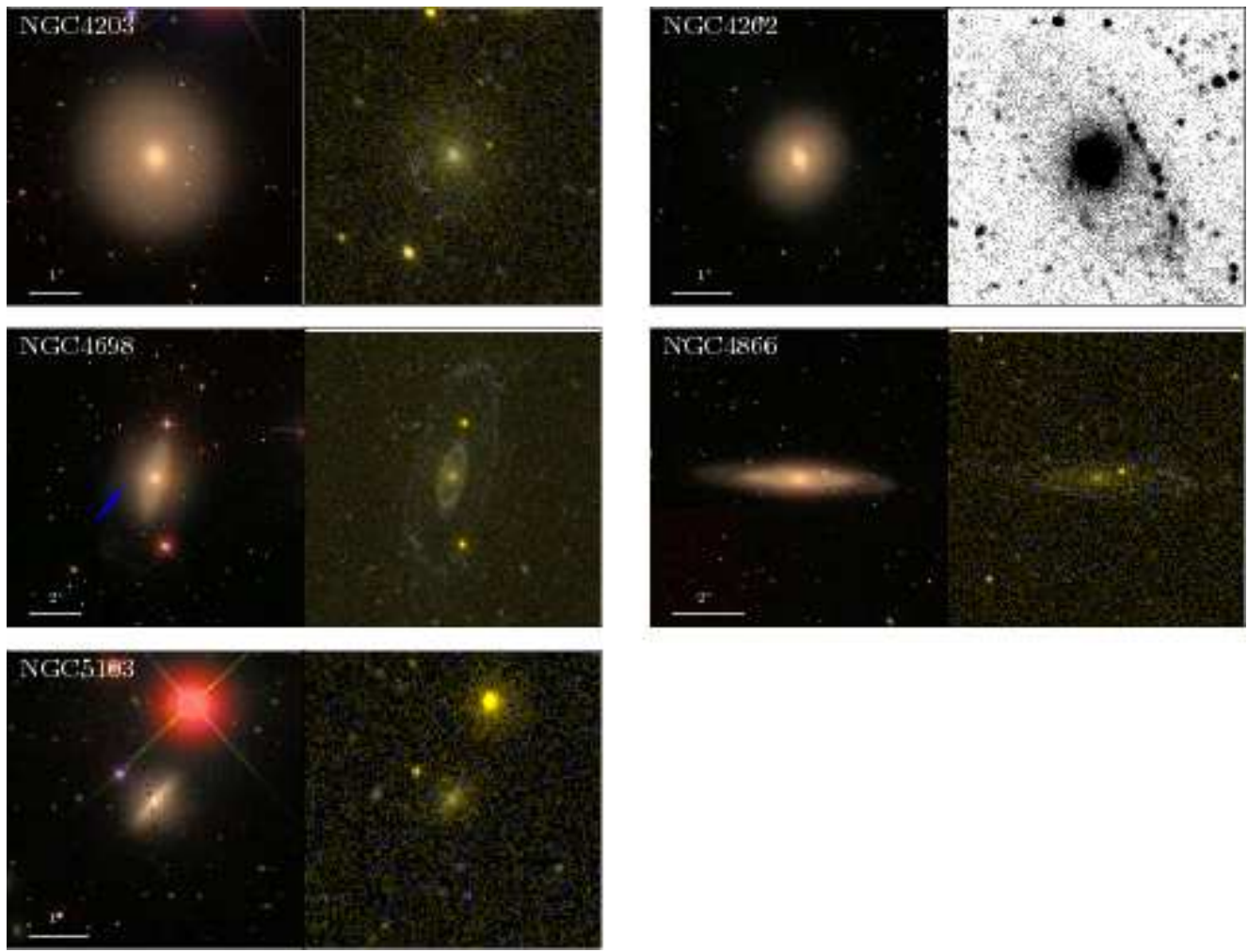}
\caption{\label{FCred} Same as Fig.~\ref{FCtrans} for \hi-normal red-sequence galaxies. We note that 
NGC4262 and NGC4698 go back to the transition region if the outer UV rings are included in the estimate 
of the UV flux.}
\end{figure*}

The presence of UV rings becomes more intriguing when \hi-normal galaxies in the `NUV red sequence' are 
also taken into account (5 galaxies in total, see Table~\ref{hirich}, Fig.~\ref{FCred} and Appendix A for 
a description of these objects).
The four galaxies with clear star formation activity (NGC4203, NGC4262, NGC4698, NGC4866) have star-forming 
regions mainly arranged in structures which are suggestive of one or multiple rings (see Fig.~\ref{FCred}).
In the case of NGC4203 and NGC4698, \hi~maps reveal a morphology similar to the one observed in \hi-normal 
transition objects with extended low surface density \hi~disks and peaks of column density in correspondence 
of UV star-forming regions. 
The only known exception is represented by NGC4262, where the \hi~is mainly segregated in the UV star forming ring.
Interestingly, NGC4262 and NGC4698 could be immediately reclassified as transition region galaxies if the 
outer UV rings are included in the estimate of the UV flux. 
All this observational evidence strongly suggests that the evolutionary paths leading these objects to 
the transition region have been significantly different from the one followed by \hi-deficient galaxies.

\subsubsection{Past and future evolution in the colour-mass diagram}
The main piece of information necessary to understand the recent evolutionary 
history of these unusual `gas-rich' systems, is the origin of their gas reservoir.
Has the gas an external origin (e.g, accretion, infall, merging, shells, etc.) or was it always part of 
the galactic halo but has not been efficiently converted into stars?
The origin of the gas can in fact not only provide clues on the past history of these objects (i.e., 
whether they are really migrating from the blue to the red sequence) but also may help in predicting 
their future evolution in the colour-mass diagram.
In the following, we combine the information available for HI-normal transition and red sequence galaxies to 
determine whether these systems have recently left the blue sequence after a quenching episode or 
are migrating back from the red sequence to the transition region thanks to a recent accretion event.

As discussed in Appendix~A, many galaxies in our sample show direct or indirect evidence of 
past gas accretion/infall events (e.g, warps, counter-rotating or decoupled components, stellar shells). 
Among the best candidates for an external origin of all the \hi~observed, there are NGC4262 and NGC4203 in 
the red sequence and NGC3619 in the transition region \citep{vandriel91}.
However, it is interesting to note that the acquisition mechanisms (and therefore the evolution) of the 
three systems is likely to be different. 
In the case of NGC3619, we are likely witnessing a minor merger with a gas-rich satellite. 
The \hi~is segregated well within the optical radius, roughly coinciding with the star forming disk observed in UV, 
suggesting that a satellite has sunk into the center initiating an episode of star formation.
Given that the stellar populations have ages and metallicities typical of unperturbed ellipticals \citep{howell06}, the most 
plausible scenario is that NGC3619 has left the red sequence after the merging event.
Interestingly, at the current SFR ($\sim$0.1 M$_{\odot}$ yr$^{-1}$), the amount of atomic 
hydrogen present within the optical disk ($\sim$7$\times$10$^{8}$ M$_{\odot}$) is sufficient 
to sustain the star formation for several billion years. Thus, NGC3619 will either remain in the 
transition region for a long time or, in case of a significant increase of the SFR, may be able to 
temporarily rejoin the blue cloud in a UV-near-infrared colour magnitude diagram. 
A similar evolutionary path could also have been followed by the dust-lane early-type NGC4370. 
However, the lack of detailed \hi~maps prevent us from drawing any conclusion.

On the contrary, the infall of \hi~into NGC4203 and NGC4262 has likely followed less `violent' paths.
In NGC4262, the presence of a ring composed only of \hi~and newly formed stars is strongly suggestive of 
recent accretion, apparently ruling out that the ring has been formed from galactic material through bar instability.
An interesting possibility is that the bar could still be responsible for the peculiar configuration of the \hi, preventing 
the newly accreted gas to collapse into the center.
What remains unclear is whether the gas in the ring has been accreted from the intergalactic medium 
(as proposed in the case of polar ring galaxies; e.g., \citealp{maccio06}) or during an interaction with another 
galaxy \citep{vollmer05b}.  
As for NGC3619, the gas reservoir in the ring is sufficient to keep the galaxy in the transition region 
for several Gyr or to move it back to the blue cloud, building-up a new stellar disk/ring.
To this regard, it is tempting to consider NGC4262 the ancestor of Hoag-type objects like NGC5701, 
thus implying that these two systems may be on their way back to the UV blue cloud. 
However, at this stage it is impossible to determine whether these two systems are at 
different stages of the same evolutionary path.

A migration back to the transition region appears instead very unlikely in the case of NGC4203. 
Despite its huge \hi~reservoir, this galaxy shows only weak traces of recent star formation activity and at this 
rate the integrated colour will not be significantly affected, leaving this object in the red sequence. 
A similar scenario could also be valid for NGC5107 and NGC4886 which already are in the red sequence. However, 
additional observations are required to unveil the evolutionary history of these systems.

In summary, for at least a few cases, observations seem to suggest that \hi-normal red galaxies 
have recently acquired atomic hydrogen and have started a new cycle of star formation activity leaving, 
at least temporarily, the red sequence.\\

For other transition galaxies, this scenario appears extremely unlikely.
This is particularly the case of NGC3898, NGC4772 and NGC4698. 
These three systems have very similar properties: i.e., Sa/Sab type with a significant 
bulge component (bulge-to-total ratio $\sim$0.2-0.4; \citealp{drory07}), \hi~ 
mainly distributed in two rings, one inside and one outside the optical radius, 
corresponding to the sites of recent star formation activity.     
HST images reveal that all three galaxies harbor a classical bulge \citep{drory07}, consistent 
with a `violent' and quick bulge formation in the past through mergers or clump 
coalescence in primordial disks (e.g., \citealp{noguchi99,kormendy04,elmegreen08}). 
In addition, the presence of a decoupled core (NGC4698) or a counter-rotating gas disk (NGC4772) 
is suggestive of a more recent accretion event (e.g., minor merger) supporting 
an external origin for at least part of the \hi~in these objects. 
The preferred explanation for the properties of NGC4698 is in fact a later formation of 
the disk through the acquisition of material by a completely formed spheroid.
Thus, it would be natural to argue that we are witnessing the build-up of the disk and that these 
galaxies are gradually moving from the red  to the blue cloud, following a path consistent 
with what expected by hierarchical models \citep{baugh96,kauffmann96}.
However, a rough time-scale argument rules out this hypothesis. 
All three galaxies harbor massive stellar disks (M$_{disk}\sim$1-3$\times$10$^{10}$ M$_{\odot}$ depending 
on the mass-to-light ratio difference between bulge and disk) and a SFR of 
$\sim$0.7-2 M$_{\odot}$ yr$^{-1}$ during the last Hubble time is thus necessary to form the observed disks.    
These SFRs are more typical of blue-cloud galaxies and a factor $\sim$10 larger than 
the SFR observed in these systems ($SFR\sim$0.07-0.2 M$_{\odot}$ yr$^{-1}$).
While minor mergers have probably affected these systems, we can exclude that a great part of the 
stellar disk is composed of `accreted stars', since the whole stellar component of the 
satellite is supposed to collapse into the center, contributing to the growth of the 
bulge \citep{hopkins09}.
Although the uncertainties in the estimate of stellar masses and SFR are still quite significant, the large 
discrepancy between the observed and expected SFR suggests that the existing stellar disks 
are too massive to have been formed at the current SFR.
Of course, we cannot exclude multiple transitions from the blue to the red sequence and vice-versa \citep{birnboim07}, 
but given the low amount of observational constraints available, we prefer not to include 
such scenario in our analysis.
We thus propose that the SFR in the disks was higher in the past or, in other words, 
that these galaxies have probably migrated from the blue cloud.
A reduction in the SFR is likely due to the low \hi~column density in these systems: on 
average below the threshold for the onset of star formation.
What caused this reduction is still unclear and only more detailed theoretical models and 
simulations will help us to solve this mystery.
The same mechanism is probably behind the ring-like structures 
observed in both \hi~and UV. 
Both internal (e.g, bar instability) or external (e.g. accretion, merging) processes can be responsible 
for such features. However the absence of strong bars, the presence of decoupled/counter rotating components and the 
size-ratio of the inner and outer rings ($\sim$3.3-3.6, i.e. different from the typical 
value expected for Lindblad resonances $\sim$2.2, although not completely inconsistent; \citealp{athanassoula82,buta95}), 
favour an external mechanism behind the unusual properties of these systems.
Since all these galaxies harbour an AGN, it is natural to think about AGN-feedback. However, these are generally 
low-energetic AGNs and no direct evidence (e.g., jets) supporting this scenario has been found so far. Moreover, 
it is not clear how star-forming rings and low surface density \hi~disks can be formed via AGN-feedback.

Finally, it is important to note that, whatever the past evolutionary history of these systems, the hydrogen reservoir 
available can sustain the current star formation activity in these galaxies for at least 2-5 Gyrs 
(ignoring molecular hydrogen, helium and recycling). 
Thus, if star formation will remain as efficient as it is now, it will take a long time for these galaxies 
to reach the red sequence in a UV-optical/near-infrared colour-mass diagram. 

Unfortunately, no speculation can be made about the past evolutionary history of 
NGC4324 and NGC4378 given the lack of multiwavelength observations.

\begin{figure*}
\includegraphics[width=17.5cm]{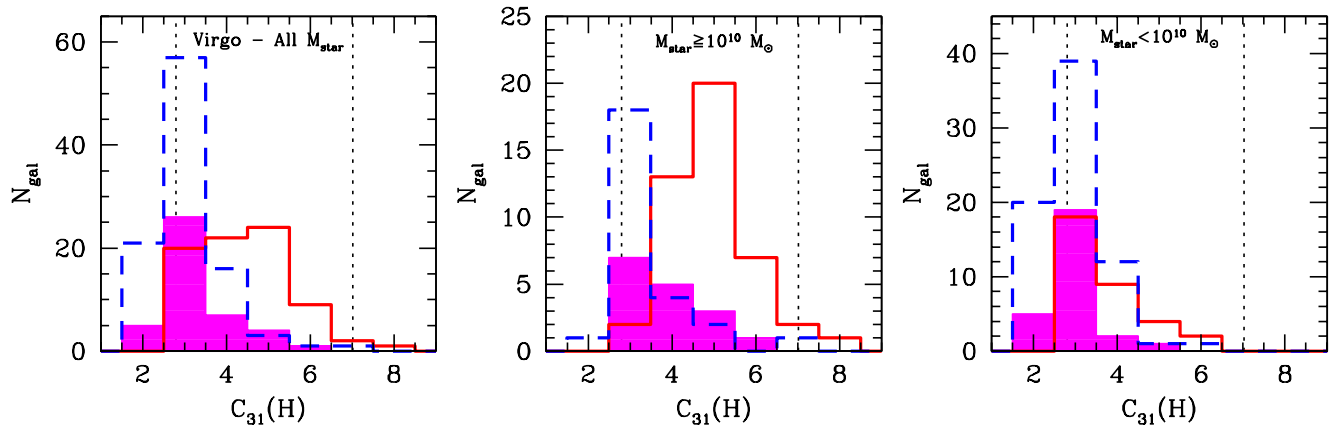}
\caption{\label{c31virgo} The distribution of the concentration index in H-band for Virgo cluster galaxies.
Galaxies are divided into blue-cloud (blue, dashed histogram), 
\hi-deficient transition (magenta) and red-sequence (red) objects. The left panel shows all the 
galaxies in our Virgo sample, while in the central and right panels only high and low stellar mass objects 
are shown respectively. The dotted lines show the expected values for an exponential and r$^{1/4}$ light
profiles.}
\end{figure*}

\section{Discussion}
Our analysis provides definitive evidence that galaxies in the transition region of the 
$NUV-H$ colour-mass diagram are a heterogeneous population.
This result strongly suggests that galaxies lying between the blue and red sequence 
have followed different evolutionary paths, not always going towards redder colours.

We remind the reader that the difference in number density between \hi-deficient and \hi-normal transition 
galaxies must be taken with a grain of salt. Although our sample is magnitude- and volume-limited, it might be 
biased towards high-density environments. Nearly half of the galaxies in our sample lies in fact within 
the Virgo cluster, which might not be a fair representation of the local universe.

Interestingly, external (i.e., environmental) mechanisms are almost always behind the 
peculiar properties of transition galaxies, both \hi-deficient and \hi-normal objects. 
Although the majority of high-mass ($M_{star}$\gs 10$^{10}$ M$_{\odot}$) transition galaxies 
harbour an AGN, feedback from accreting super-massive black holes appears not to be necessary 
to explain their properties. 
As discussed in HC09, the presence of AGNs in transition galaxies does not automatically 
imply a physical link between nuclear activity and quenching. 
Moreover, we do not find any direct observational evidence (e.g., jets, radio lobes, etc.) supporting an 
interaction between the central black hole and the galaxy's gas reservoir.

\subsection{The migration to the red sequence of \hi-deficient galaxies}
\hi-deficient transition galaxies constitute the majority of the transition 
population in our sample. By extending the analysis presented in 
HC09, we have shown that these galaxies not only are mainly found in the Virgo cluster but 
they also are the only population which is clearly migrating from the blue towards the red sequence. 
While environmental effects are certainly able to strip the gas from the 
disk, reducing the star formation in just a few hundreds million years and 
forcing the galaxy to leave the blue cloud, 
less clear is the last leg of the journey, i.e. the migration from the transition 
region to the red sequence. 
The complete suppression of the star formation requires at least a few billion years.
This `two-step' migration is more dramatic, and perhaps only visible, in a 
UV-near-infrared colour-mass diagram whereas in optical the first stripping 
event is sufficient to make the colours almost as red as an early-type galaxy.

At this stage, it is very tempting to use our data to quantify the current `migration' rate and, 
consequently, the mass accretion rate of the red sequence. 
Unfortunately, the large uncertainties in both observables and 
on the typical time-scale of the migration (at least a factor 2) make this exercise not 
very useful. For example, a quenching time of $\sim$3 Gyr would suggest that, 
at the current rate, the red sequence in our sample could have been built by the migration 
of objects from the blue cloud in a Hubble time, consistent with previous works 
(e.g., \citealp{arnouts07,martin07,schiminovich07}). 
However, at the same time, we are not able to reject scenarii in which either the observed rate 
is able to build-up the red sequence in half a Hubble time or the observed migration is not 
able to explain the growth of the quiescent galaxy population in the last 13 Gyr. 
Thus, no additional constraints on the evolution of the colour-stellar mass diagram 
are imposed by the estimate of the stellar mass accretion rate observed in our sample.

A more interesting exercise is to look for any morphological transformation during the migration 
towards the red sequence. The crucial question here is whether  
the red sequence is fed with bulge dominated or disk galaxies. 
The answer is clear from Fig.~\ref{c31virgo}, where we compare the distribution of the 
concentration index in H-band for galaxies in the Virgo cluster. 
In the transition region, we show only Virgo \hi-deficient galaxies.
Overall (Fig.~\ref{c31virgo}, left panel), \hi-deficient transition galaxies have a concentration 
index much more similar to blue than red-sequence systems.
However, such difference is only evident at stellar masses higher than $\sim$10$^{10}$ M$_{star}$ 
(Fig.~\ref{c31virgo}, central panel) whereas for smaller galaxies (right panel) the 
distribution of $C_{31}(H)$ does not significantly vary across the whole range of colours,  
reflecting the fact that dwarf ellipticals have exponential light profiles like dwarf irregulars 
(e.g.,\citealp{binggeli91}).  
This result implies that, while \hi-deficient transition 
galaxies are likely the progenitors of cluster low-mass red objects (see also \citealp{haines08,dEale}), 
this is not completely 
true at high stellar masses. 
This is additionally supported by the fact that the vast majority of high-mass transition galaxies 
are early-type spirals and almost no ellipticals are present (Fig~\ref{typedistr} and \ref{trdistr}).  
The vast majority of galaxies in the process of reaching the red sequence are thus disk systems, 
significantly different from ellipticals or bulge-dominated galaxies characterizing 
high-mass, quiescent objects at low redshift.

Since a significant fraction of \hi-deficient transition 
galaxies appears to have recently infalled into the center of Virgo and will likely 
spend a few Gyr in the transition region (see \S~\ref{hidef}), it may be possible 
that a morphological transformation still takes place before reaching the red sequence.
Although we cannot completely exclude this scenario, we consider it unlikely. 
Given the long time required to halt the star formation, galaxies with increased bulge component 
should be present in our sample. Moreover, since gas stripping appears not to significantly 
increase the bulge component in early type galaxies (e.g., \citealp{n4569}), 
an additional environmental effect (different from the one responsible for the 
quenching of the star formation) must be invoked.

Thus, the fact that transition galaxies are not morphologically transformed before 
reaching the red sequence may have two different implications: either 1) galaxies are morphologically transformed 
once already in the red sequence, or 2) the mechanism controlling the accretion of 
stellar mass into the red sequence at $z\sim$0 is not the one responsible for the 
creation of the red sequence in the first place.
Although it is possible that the bulge component is enhanced in some 
red galaxies via gravitational interactions, such scenario seems 
unlikely to explain the growth of the red sequence.
Firstly, the mechanism responsible for the morphological transformation should be 
efficient only on giants and not on dwarf systems, which seems inconsistent with what known 
about environmental effects \citep{review}. 
Secondly, the major mergers required to significantly increase the bulge component are extremely 
rare in today's clusters of galaxies.
Thirdly, the presence of red-sequence galaxies in isolation (e.g., HC09) implies that 
star formation has been suppressed also outside clusters of galaxies.
Although our sample could be biased against isolated objects, the fact that we do not find a 
field population which is clearly migrating from the blue sequence 
may suggest that in low density environments the red sequence has been mainly 
populated in the past.
Finally, if the morphological transformation takes place in the 
red sequence, the number of bulge-dominated/elliptical galaxies should decrease at 
increasing redshift, which does not seem to be the case \citep{postman05,desai07}.
Thus, the most favourite scenario emerging from our analysis is that the red sequence is 
currently accreting mass, at all masses, mainly via disk galaxies. 
No significant structural modification takes place during the journey from the blue cloud 
to the red sequence. 
The main process responsible for the suppression of the star formation in nearby galaxies is 
thus not the same responsible for the formation of the red sequence at high redshift.

\begin{figure}
\includegraphics[width=8.5cm]{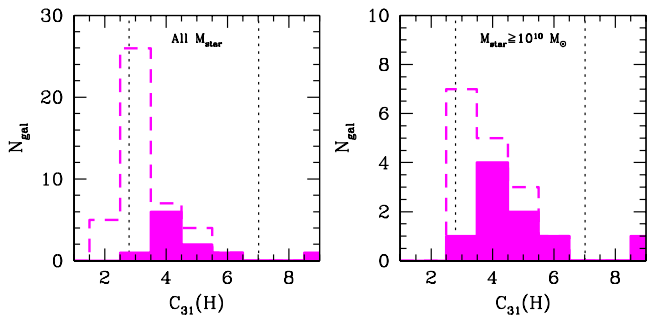}
\caption{\label{c31gasrich}. The distribution of the H-band concentration index for Virgo \hi-deficient 
(dashed) and \hi-normal transition galaxies. All galaxies and galaxies with $M_{star}\geq10^{10}$ 
M$_{\odot}$ are shown in the left and right panel respectively.}
\end{figure}

\subsection{The \hi-normal side of the transition region}
While in the case of \hi-deficient objects it seems plausible to associate transition galaxies 
with objects that are migrating from the blue to red sequence, this is not 
always the case for \hi-normal transition galaxies. 
The discovery of such systems is probably the most exciting result of this work.
They represents  $\sim$12\% of our transition galaxy population, and are only found 
at high stellar masses ($M_{star}>$10$^{10}$ M$_{star}$).

As shown in Fig.~\ref{c31gasrich}, these galaxies are mainly disks with a significant bulge 
component {\bf (they are in fact all Sa or S0 galaxies)} and, despite the low number statistics, 
it seems clear that, , contrary to the \hi-deficient population, in this family we find very 
few `disk-only' galaxies. 
Despite their different properties, a great fraction of these galaxies show 
active star-forming regions and \hi~segregated in one or multiple ring-like structures.
The picture emerging from our analysis is quite complex and exciting at the same time, 
showing that the transition region can be fed with galaxies from both sequences.  

Red sequence galaxies can acquire new gas supply and restart 
their star formation activity, as predicted by cosmological simulations. 
However, it is unlikely that the red galaxies in our sample will re-build a significant 
stellar disk (see also \citealp{hau08}). Merging, accretion of gas-rich satellites and 
exchange of material during close encounters are among the likely responsible for this rejuvenation.
Such processes are more frequent in low density environments, where red-sequence galaxies are rarer, 
perhaps reducing the chances to observe such phenomenon.
 
When a suppression of the star formation is the most likely scenario to explain \hi-normal transition 
galaxies, the process behind such migration is still unclear.
In this case, star formation must be reduced by making the \hi~stable 
against fragmentation (e.g., by decreasing the \hi~column density below 
the threshold for star formation) and not via \hi~stripping as observed in \hi-deficient objects.
Starvation \citep{larson80} by removing any extended gaseous halo surrounding 
the galaxy, preventing further infall, could be a possibility. This 
would imply a longer time-scale (several Gyr) for the migration 
from the blue cloud \citep{n4569,dEale} than the one observed in the case of \hi~stripping 
(a few hundreds million years).  
However, the evidence for accretion/interaction in some of these objects may suggest that 
starvation, if efficient, is not the only mechanism at work.

\cite{martig09} have recently proposed a `morphological quenching' to 
explain the origin of gas-rich bulge-dominated objects. 
The idea behind this mechanism is that the presence of a bulge could inhibit the collapse 
of a gas disk. However, the `morphological quenching' appears only to be effective when 
the disk stellar component is negligible, which is not the case for the majority 
of the systems in our sample.

Finally, the fact that the majority of these objects show some level of nuclear activity, might indicate 
a link between AGN activity and their position in the colour stellar mass diagram. 
However, although we cannot exclude that recent accretion events may have triggered the AGN, we do not find any direct 
observational evidence suggesting that AGN-feedback is playing a significant role in the recent star formation history of 
these objects (see also HC09).

Thus, the past evolutionary history of \hi-normal transition galaxies has still to be unravelled. 
Whatever is the path followed to get to the transition region, \hi-normal systems currently 
have enough fuel to sustain star formation at the current rate for almost 
another Hubble time. This is an unexpected result, implying that such systems could 
remain in the transition region for a very long time and that the colour range 4.5$<NUV-H<$6 mag 
does not automatically correspond to a snapshot in the star formation history of 
nearby galaxies.

\section{Summary \& Conclusion}
In this paper, we have combined UV, \hi~and near-infrared observations to investigate 
the properties of local transition galaxies.
Our main results are as follows.\\

i) We confirm that the reddening of the galaxy colour is accompanied by a decrease in the 
\hi~gas-fraction. However we show that, while in the blue cloud colour and gas-fraction 
are tightly correlated, in the transition region and red sequence such correlation is more 
scattered. Transition region galaxies can thus be divided into two main families according to 
their \hi-content.\\

ii) \hi-deficient transition objects reside in high-density environments and environmental 
effects are the likely cause for the loss of gas. Star formation is rapidly suppressed after the gas 
stripping and the transition region is reached in less than a billion years.
However, once in the transition region, massive galaxies are still forming stars 
and will not immediately reach the red sequence. A subsequent quenching is thus required to 
reach the red sequence, implying that the total migration time-scale is, at least, a few billion 
years. \\

iii) Although \hi-deficient galaxies represents the bulk of the migrating population 
in our sample, they {\bf are apparently not responsible} for the formation of the red sequence in the first place. 
The gas stripping is in fact `polluting' the red sequence mainly with disk galaxies and no morphological 
transformation is observed during the quenching of the star formation. \\

iv) Contrary to \hi-deficient systems, \hi-normal transition galaxies ($\sim$12\% of the whole transition 
population in our sample) represent an heterogeneous population 
of objects, at least two of which are probably following the inverse path, migrating back from the red sequence.
The high hydrogen content in these systems is, at least in some cases, due to external 
accretion/interaction events. The detailed evolution of these objects is still unclear, but it is likely 
that they will remain in the transition region for several billion years. Thus, a connection 
between transition region and migration from the blue to the red sequence may not always be true.\\

The discovery of a \hi-normal population in the transition region has only been possible thanks 
to the combination of UV imaging (necessary to properly separate blue and red sequence at all 
masses) and \hi~single-dish observations (to quantify the atomic hydrogen content).
The results here presented have inevitably lead to several questions we have not been able to answer in the 
current work. Is accretion really playing a crucial role in the evolution of these objects? 
Which mechanism suppressed star formation without removing the gas?
How frequent are such objects? etc. 
To be able to answer these questions, both detailed theoretical and observational studies of individual objects and 
statistical investigations of larger samples are mandatory.
Luckily, we are currently in an exciting time for UV and 21 cm astronomy. 
Particularly promising for this topic is the {\it GALEX Arecibo SDSS Survey} 
(GASS, \citealp{catinella08}) which is currently 
making a census of the \hi~content in a complete sample of $\sim$1000 massive galaxies ($M_{star}\geq$10$^{10}$ M$_{\odot}$), 
ideal to investigate the role of gas and accretion on the evolution 
of transition galaxies.
Thus, we are probably not too far from being able to unravel some of the mysteries 
still surrounding \hi-normal galaxies in the transition region.

\section*{Acknowledgments}
We are greatly indebted to Alessandro Boselli for his support and encouragement in carrying 
out this analysis and for providing part of the data before publication.
We wish to thank Alessandro Boselli, Barbara Catinella, Jonathan Davies and Rory Smith 
for useful discussions and helpful comments on the manuscript.  
We thank the anonymous referee for useful comments which helped to improve this paper.
LC thanks the hospitality of the Max Planck Institute for Astrophysics where part of 
this paper was written.  
LC and TMH are supported by the UK Science and Technology Facilities Council.

This publication makes use of data products from the Two Micron All Sky Survey, 
which is a joint project of the University of Massachusetts and the Infrared Processing and Analysis 
Center/California Institute of Technology, funded by the National Aeronautics and Space 
Administration and the National Science Foundation, from the GALEX mission, developed in cooperation with the  
Centre National d'Etudes Spatiales of France and the Korean Ministry  
of Science and Technology, from the NASA/IPAC Extragalactic Database (NED) which is operated by the 
Jet Propulsion Laboratory, California Institute of Technology, under contract with the National 
Aeronautics and Space Administration and from the GOLDMine data base.

This research has made extensive use of NASA's Astrophysics Data System and of the astro-ph preprint 
aechive at http://arXiv.org.

\section*{Appendix A: Notes on individual objects}

\subsection*{\hi-normal transition galaxies}
{\bf NGC3619} is a S0 galaxy, member of a group composed of at least five systems \citep{garcia03}.
The \hi~distribution is asymmetric and much concentrated in the center \citep{vandriel89}, 
where star formation is taking place and dust lanes are clearly visible.
The extent of the \hi~roughly matches the UV star-forming disk, i.e., $\sim$half the optical radius.
In optical, the galaxy shows prominent outer stellar shells, suggesting a recent minor-merging/accretion event.
Numerical simulations predict that the stars from a satellite make a system of shells 
several 10$^{8}$ yr after the end of the merging event which can last for more than 1 Gyr \citep{kojima97}. 
The accretion scenario is also supported by the spectroscopic analysis performed by \cite{howell06}.
He finds that the age, metallicity, and $\alpha$-enhancement ratios of NGC3619 are consistent with those of a representative 
sample of unperturbed nearby elliptical galaxies, contrary to what expected in the case of a recent major merger and 
supporting a minor merger with a gas-rich dwarf satellite.
Thus, all the observational evidence is consistent with an external origin for the atomic hydrogen and 
residual star formation observed in NGC3619.  

\noindent
{\bf NGC3898} is a Sa galaxy in the Ursa major group. In the Hubble Atlas it is described as the prototype of Sa 
galaxy: i.e., with a prominent bulge component and multiple spiral arms. 
HST observations reveal that the bulge is a `classical bulge' \citep{drory07}, 
consistent with a major merger/accretion event in the past. 
UV observations show that active star formation is mainly segregated in a ring well 
within the optical radius, although star-forming regions are also found at the edge of the optical disk.
The atomic hydrogen is distributed in a low column density ($\sim$1-2 M$_{\odot}$ pc$^{-2}$) disk extending up 
to $\sim$4 times the radius containing 80\% of the B-band light \citep{noordermeer05}.
The highest \hi~column density is observed in correspondence with the UV star-forming regions, suggesting that only 
in these few regions the \hi~can condensate into molecular hydrogen and initiate the star formation cycle \citep{krumholz09}.
The \hi~disk appears to be warped in the inner parts and shows some `wiggles' with an amplitude of 
30-50 \kms \citep{noordermeer07}.
Although H$\alpha$ long-slit spectroscopy 
\citep{pignatelli01,vega01} and 21 cm \hi~line interferometry \citep{noordermeer05,noordermeer07}  
suggest that both ionized and atomic hydrogen have a regular velocity field and a smooth distribution, 
\cite{noordermeer05} suggest that this galaxy may have experienced a recent interaction/accretion event.
However, the basis for this interpretation is not completely clear.

\noindent
{\bf NGC4324} is a Sa galaxy belonging to the Virgo S cloud \citep{goldmine}.
Star formation is segregated in a ring, also visible in SDSS images, distinct from the central bulge component. 
Although \hi~maps are not available, \cite{duprie96} detect \hi~up to $\sim$2 optical diameters suggesting that 
the atomic hydrogen is not only concentrated in the star-forming ring.

\noindent
{\bf NGC4370} is an early type galaxy member of the Virgo B cloud \citep{goldmine} forming a non-interacting 
pair with NGC4365. Its detailed morphological classification is rather uncertain, varying from Sa \citep{vcc} to elliptical 
\citep{bertola88}. This is mainly due to the presence of a prominent dust lane along the equatorial 
plane of the galaxy. 
Although active star-forming regions are not clearly evident from UV and H$\alpha$ images, we cannot exclude that 
we are observing an analogous of the Sombrero galaxy (NGC4594) completely edge-on.
Excluding the prominent dust lane, no other evidence for interaction/accretion is found and the gas and stars in the 
galaxy are co-rotating \citep{bertola88}. 
Surprisingly, \cite{bertola88} interpret the co-rotation in NGC4370 as a sign of possible accretion, 
suggesting that the gas present in this galaxy has an external origin.
Although the basis for this interpretation may be debatable, 
the same conclusion has been recently reached by \cite{patil09}, who investigated the dust content in NGC4370.
They propose that the amount of dust observed in the galaxy can not have been formed in situ and that 
at least part of the material must have been accreted during an interaction event.
An external origin for the atomic hydrogen content in this galaxy is thus not excluded, as observed in  
other \hi-rich dust-lane elliptical galaxies \citep{oosterloo02}.

\noindent
{\bf NGC4378} is a Sa galaxy belonging to the Virgo S cloud. It has a prominent nuclear bulge 
surrounded by a low surface brightness disk which contains a single, tightly-wound, spiral arm. 
Given the low surface brightness of the disk component, \cite{vandberg76} classified this object as an early-type 
anemic spiral, although we now know that its atomic hydrogen content is not significantly different from the one in high 
surface brightness disks. UV star formation is clearly visible across all the disk.  
Numerical simulations \citep{byrd94} suggest that the unusual one-armed spiral structure in NGC4378 is 
in reality an impulsive trailing arm created by the recent passage ($\sim$200 Myr) of a small companion 
($\sim$1/30 the galaxy's mass). 
Since the satellite is supposed to survive the interaction, it seems unlikely that a significant fraction 
of the \hi~present in NGC4378 has been recently accreted.
The low SSFR of NGC4378 may therefore have an internal origin. 
By studying a sample of nearby galaxies, \cite{seigar05} recently found a correlation between 
the average shear rate in disks and their SSFR, suggesting the existence of a 
shear-threshold above which star formation is inhibited. 
Interestingly, NGC4378 has a shear rate (0.69$\pm$0.03) just below the threshold value (0.70$\pm$0.09) 
suggested by \cite{seigar05}.
If this is the case, the internal dynamical properties of NGC4378 may be responsible for the low star 
formation efficiency observed in this object. 

\noindent
{\bf NGC4772} is a Sa galaxy in the Virgo S cloud. 
Recently, \cite{haynes00} carried out a detailed multiwavelength analysis 
of this object, showing that it is probably the result of a minor merger event.
The main observational evidence supporting such scenario is: 1) the presence of dynamically decoupled 
central gas and stellar components, in particular in correspondence of the star forming ring visible in UV (Fig.~\ref{FCtrans}) and 
H$\alpha$, 2) the unusual \hi~distribution, which is segregated in two concentric rings, the inner one where star formation 
is taking place and 3) the presence of a warp in the outer \hi~ring. 
The most likely scenario is that NGC4772 is at the end stage of a prograde merger in which the transfer of angular 
momentum leads to an outward spread of the disk \citep{quinn93,haynes00}.

\noindent
{\bf NGC5701} is the bluest object in the transition region. 
We classified this galaxy as an Sa (NED gives S0/a). 
However its peculiar morphology, characterized by a central bulge surrounded by a faint and apparently detached 
star-forming ring, makes NGC5701 a possible Hoag-type object \citep{hoag50}.
In this case the central condensation is a bar, not a `bulge or elliptical' \citep{hoag50} and \cite{gadotti03} dubbed this 
object `a barred galaxy without a disk'.
The origin of Hoag-type objects is still a puzzle, but two different scenarii seem to be the most commonly accepted: 
a) accretion of small satellites \citep{schweizer87}, or b) a strong bar instability \citep{brosh85}, which destroys 
the disk and builds up a ring (e.g. \citealp{schwarz84,byrd94b}). 
The lack of detailed dynamical and spectroscopic data prevent us from driving a conclusion on the origin of 
the ring in NGC5701, but it is quite likely that the bar is playing an important role either destroying the disk 
or preventing the accreted gas from infalling into the center.

\subsection*{\hi-normal red-sequence galaxies}
\noindent
{\bf NGC4203} is an, apparently isolated, barred S0 galaxy. 
UV images show low level star-formation at least in the south-east part of the galaxy, corresponding 
to the outer edge of the warm-dust spiral structure detected at 8$\mu$m in the galaxy center \citep{pahre04}.    
Despite its low star formation activity, NGC4203 has a large reservoir of atomic hydrogen, distributed in 
a peculiar, filamentary structure and extended up to $\sim$2.2 optical radii \citep{vandriel88,noordermeer05}.
The atomic gas has a general sense of motion around the center, but it is clearly not on regular circular orbits 
\citep{noordermeer05}. Again, the \hi~is mainly found in two ring structures: an inner ring (in correspondence 
with the UV features) and an outer ring which is likely rotating on a different plane \citep{vandriel88}. 
The peculiar dynamics of NGC4203 is thus suggestive of an accretion event \citep{vandriel88}.
The \hi~disk of NGC4203 is also one of the nearest damped Lyman alpha systems. \cite{miller99} 
found that the metallicity of the gas is significantly lower than the one observed in galaxies of the same luminosity.
If confirmed, this would represent a strong observational evidence supporting the external origin for the \hi.

\noindent
{\bf NGC4262} is a barred S0 galaxy belonging to the Virgo main cloud A. 
No star formation is observed within the optical radius. However, UV images reveal the presence of 
a UV star-forming ring extended up to $\sim$2.5 optical radii. Although faint UV emission is marginally 
visible all across the ring, 
great part of the star formation is taking place in the south-west. 
This roughly coincides with the peak in \hi~distribution as revealed in the maps presented by \cite{krumm85} and 
\cite{minchin07}. Moreover, all the \hi~is segregated into a ring and no low surface density \hi~
disk is detected. \cite{krumm85} and \cite{vandriel91} suggested that the ring has probably an external origin 
and more recently \cite{vollmer05b} tentatively proposed that the \hi~could have been accreted during an interaction 
between NGC4262 and NGC4254. 
Being a barred system plus a ring, NGC4262 shows some similarities with the 
Hoag-type object NGC5701. If this is the case, NGC4262 is probably at an earlier evolutionary stage, since its ring 
is only visible in UV whereas the ring in NGC5701 is evident also in optical, suggesting an older age for 
the stellar populations.
We note that, if we integrate the UV light up to the extent of the ring, NGC4262 would move from the red sequence to 
the transition region.  

\noindent
{\bf NGC4698} is a bulge-dominated Sa galaxy in the Virgo E cloud. 
The Carnegie Atlas \citep{sandage94} describes it as an elliptical-like bulge in which there is no evidence of
recent star formation or spiral structure, plus a low-surface-brightness disk with spiral arms become 
prominent only in the outer part of the disk.
HST observations reveal that the bulge is a `classical bulge' \citep{drory07}, consistent with a 
major merger/accretion event in the past. 
Once again, UV images reveal that star formation is segregated into two `almost perfect' rings: 
one within and the other outside the optical radius.
Interestingly, the two rings are not concentric: the inner ring is centered on the nucleus, 
whereas the outer one is offset $\sim$ 20 arcsec to the north. 
The peculiarity of NGC4698 becomes more evident when the dynamical properties of the bulge and disk 
component are compared. \cite{bertola99} showed that the rotation axis of the disk and of the bulge are almost 
perpendicular.
They interpret this unusual geometry as an evidence of a later formation of the disk through the acquisition of 
material by a completely formed spheroid.
Also in this case, the galaxy moves back to the transition region if the outer ring is included in the 
estimate of the UV flux.

\noindent
{\bf NGC4866} is an isolated, highly inclined, S0 galaxy (although \citealp{sandage94} classify it as Sa).
Very few data are available in the literature, and the only unusual feature is the presence, once again, of a UV star-forming 
ring at the edge of the optical disk. 

\noindent
{\bf NGC5103} is an isolated Sab galaxy. As for NGC3414, no recent star formation activity is clearly visible in the UV image 
and the galaxy resembles an edge-on polar ring/disk system: \cite{whitmore90} classifies NGC5103 as possible 
candidate for polar-ring galaxy. No high-resolution \hi~maps are available for this object.

Finally, a visual inspection of galaxies in the red sequence reveals the presence of two other objects with UV rings: 
NGC3945 and NGC4643.
Unfortunately, no \hi~observations are available and they are thus excluded from the discussion in the text.

\end{document}